\documentclass[aps,prd,twocolumn,preprintnumbers,nofootinbib,superscriptaddress]{revtex4}

\usepackage{color}
\usepackage{amsmath}
\usepackage{amssymb}
\usepackage{bm}

\def\beq{\begin{equation}}
\def\eeq{\end{equation}}
\def\bea{\begin{eqnarray}}
\def\eea{\end{eqnarray}}
\def \lsim{\mathrel{\vcenter
     {\hbox{$<$}\nointerlineskip\hbox{$\sim$}}}}

\def\a{\alpha}

\def\g{\gamma}

\begin{document}

\title{Axion absorption and the spin temperature of primordial hydrogen}

\author{Adrien Auriol}
\affiliation{LUPM, CNRS, Universit\'e Montpellier,
Place Eugene Bataillon, F-34095 Montpellier, Cedex 5, France}

\author{Sacha Davidson}
\affiliation{LUPM, CNRS, Universit\'e Montpellier,
Place Eugene Bataillon, F-34095 Montpellier, Cedex 5, France}

\author{Georg Raffelt}
\affiliation{Max-Planck-Institut f\"ur Physik
  (Werner-Heisenberg-Institut), F\"ohringer Ring 6, 80805 M\"unchen, Germany}

\begin{abstract}
An absorption dip in the spectrum of the cosmic microwave background
observed by the EDGES experiment suggests an unexplained reduction of
the hydrogen spin temperature at cosmic redshift $z\sim17$. The mass
of dark-matter axions could correspond to the hyperfine splitting
$E_{10}=5.9~\mu{\rm eV}$ between the triplet (H$_1$) and singlet
(H$_0$) state. { 
  We calculate 
 the rate for  $a+{\rm H}_0\leftrightarrow{\rm
   H}_1$ in two ways,    and find that it is orders
of magnitude smaller than the CMB-mediated transition rate, so irrelevant. 
As a result,   this process cannot
be used to rule in or out dark matter axions with $ m_a\sim E_{10}$.
The axion rate  nonetheless has interesting features,
for example,  on balance  it   heats the spin
temperature, and  the axion couplings to
 protons and electrons contribute on equal footing. } 
\end{abstract}

\maketitle


\section{Introduction}
\label{intro}

The EDGES experiment \cite{EDGE} has seen an absorption line in the
Cosmic Microwave Background (CMB) spectrum that corresponds to 21~cm
at cosmic redshift $z\sim 17$ and thus to the hyperfine splitting of
the hydrogen ground state.  This feature is expected, prior to
reionisation, but it is too deep, suggesting that the upper (triplet)
level~H$_1$ is underpopulated relative to the lower (singlet)
level~H$_0$.  Equivalently, the spin temperature of hydrogen is cooler
than expected.  Various explanations of this observation have been
proposed; some involve cooling the spin temperature via interactions
with dark matter \cite{Tashiro:2014tsa,Munoz:2015bca,
  Barkana:2018lgd, Fialkov:2018xre,
  Barkana:2018qrx, Berlin:2018sjs, LM, gravtherm1, gravtherm2,
  otheraxion}, others add a radio background
\cite{Feng:2018rje,Ewall-Wice:2018bzf} or
heat the CMB photons in the appropriate
frequency range \cite{PPRU,Moroi,FalkowskiPetraki}. The EDGES
observation has also been used to constrain the properties of
dark matter \cite{DMconstraints1, DMconstraints2, DMconstraints3,
  DMconstraints4, DMconstraints5, DMconstraints6, DMconstraints7} and
axion-like particles \cite{Mirizzi}.

One idea on how axions might cool the hydrogen spin temperature relies
on gravitational thermalisation among axions and with baryons
\cite{gravtherm1, gravtherm2}. Axions are cold dark matter because
they are born non-thermally as low-momentum classical field
oscillations. Their momentum distribution depends on the exact
early-universe scenario and may include the formation of bound objects
such as axion miniclusters. Unbound axions interact extremely weakly,
yet it has been argued that they can equilibrate gravitationally where
the weakness of the interaction is enhanced by coherence over
large-scale density fluctuations. One conceptual difficulty of this
scenario is that a kinetic treatment of relaxation is not
straightforward because one is in the condensed regime.
Gravitational axion thermalisation has been controversially
discussed and certainly is not a trivial issue.

We are here interested in another way in which axions would affect the
spin temperature, i.e., by direct absorption or emission $a+{\rm
  H}_0\leftrightarrow{\rm H}_1$ \cite{LM}. The hyperfine splitting
energy $E_{10}=5.9~\mu{\rm eV}$ corresponds to an axion mass where
these particles could well be most or all of dark matter. Therefore,
if the oscillating axion field were on resonance with the  hyperfine
splitting, large effects are conceivable\footnote{Spin-flip transitions
in other atoms, which have similar properties to the
hydrogen hyperfine transition,
have also been considered as a means to detect galactic axions
\cite{Sikivie:2014lha}.}.  Assuming $m_a=5.9~\mu$eV
and noting that the dark matter density is about four times that of hydrogen,
there would be around $10^{15}$ axions for every H
atom. So less than one in $10^{15}$ axions needs to interact to have
a strong effect. On the other hand, axion interaction rates are
extremely small, so some significant enhancement is needed to
achieve a large effect.

The purpose of this short note is to study the rate for the processes
$a+{\rm H}_0\leftrightarrow{\rm H}_1$ for the conditions pertaining at
Cosmic Dawn. One simple result, applicable if  these interactions
are in equilibrium, is that the large axion
occupation numbers would  imply that  the two atomic levels
would be nearly equally populated to balance the forward and backward
rates. In other words, the equilibrium spin temperature would be
almost infinite even though axions are cold in the sense of their
kinetic distribution.

Therefore, this process cannot explain the EDGES observation, but one
could turn it around and exclude axion dark matter in a narrow mass
range. However, the rate for the absorption process is very small and
there does not seem to exist any large enhancement factor.  We arrive
at this conclusion in two ways. One is a simple kinetic calculation
based on a squared transition matrix element and the phase-space
occupation factors for the axion field modes. The other builds on
interpreting dark-matter axions as a classical field oscillation. As
such the problem is analogous to a two-level atomic system interacting
with a laser beam on or near resonance. Both calculations arrive at
the same result as expected.

We mention in passing a curious detail about the axion hyperfine
transition: it depends symmetrically on the axion interaction
with protons and electrons. This is unlike typical
atomic transitions (``axio-electric effect'') which are dominated by
electrons \cite{Redondo, PRV, Detal}. In our case, hadronic axion
models provide a particularly generic transition rate whereas in
non-hadronic models it depends on the relative electron and proton
coupling strengths.

The rest of this short note is devoted to working out
these arguments in detail.
Section \ref{sec:rev} briefly reviews the
hyperfine splitting and axion dark matter.
In Sec.~\ref{sec:QFT}, we estimate in   Field Theory the
rate at which dark matter axions could  induce
hyperfine transitions,  by interacting
with either the electron or proton
of the hydrogen atom. The problem  is
analogous to  the quantum mechanical calculation
of electrons jumping  up and down  levels
in the presence of a classical  electromagnetic wave,
so in Sec.~\ref{sec:QM} we match our
problem onto the well-known solution
of  the quantum mechanical problem, given
in the textbook of Cohen-Tannoudji {\it et al.} \cite{CT}.
The final discussion, Sec.~\ref{sec:disc},
compares our rate calculation to the literature
\cite{LM,Redondo,PRV,Detal}, and discusses whether this process could heat or
cool the spin temperature.

\section{Review}
\label{sec:rev}

\subsection{Hyperfine Transition in Primordial Hydrogen}
\label{sec:hyperfine}

The 21 cm line corresponds to
the hyperfine splitting  of
the $1s$ state of the H atom, due to the interaction
of the electron spin $\vec{S}_e$ with the magnetic field
induced by the proton spin  $\vec{S}_p $
(see Refs.~\cite{CT,BjD} for an introduction).
This interaction
is proportional to the magnetic moments of the
two fermions, and,  integrated over the
probability distribution of the $1s$ state,
gives a contact interaction proportional to
\begin{equation}
  \vec{S}_p  \cdot \vec{S}_e = \frac{1}{2}
  \bigl(|\vec{J}|^2  - |\vec{S}_p|^2 -|\vec{S}_e|^2\bigr),
\end{equation}
where the sum of the  spins is the total angular momentum $J$.
(Recall the $1s$ state has
no orbital angular  momentum.) So
the energy  eigenstates are the eigenstates  of  \smash{$\vec{J}$}:
the $J=1$ triplet state with $M \in\{-1,0,1\}$ and the
$J=0$ singlet state.

The energy splitting between the singlet and triplet states
can be expressed as a numerical factor multiplying
the magnetic moments of the electron and proton, and
the square of the $1s$ wavefunction at the origin \cite{BjD}:
\begin{equation}
E_{10}
\simeq \frac{4}{3} \frac{g_e e}{2m_e}  \frac{g_pe}{2m_p}
\frac{\alpha^3 m_e^3}{\pi}\,.
\label{E10}
\end{equation}
The value of $E_{10}$ has been measured to many significant figures. For our discussion,
$E_{10}=5.9~\mu{\rm eV}=2\pi/21~{\rm cm}=0.068~{\rm K}$ is precise enough.

The relative populations of the triplet vs.\ singlet states can be
expressed in terms of the
spin temperature $T_S$ by
\beq
\frac{n_1}{n_0} = 3\exp(- E_{10}/T_S),
\label{TS}
\eeq
where the $n_i$ are number densities and
3 is the statistical weight of the triplet state.

The EDGES experiment looks for  an absorption line, in the cool tail
of the CMB, corresponding to the 21~cm  line of  neutral hydrogen
in the epoch prior to reionisation.
Absorption should occur because the
spin temperature of hydrogen is expected
to be cooler than the CMB, which at $z \sim 17$  is at
$T_{\rm CMB} \simeq (1+z)\,2.7~{\rm K} \simeq 48~{\rm K}$.
For a review of 21~cm line cosmology, see Ref.~\cite{Furlanetto:2006jb}.

Three processes that  can change $T_S$
at $z\sim 17$
are \cite{Furlanetto:2006jb}
decays and photon absorptions $H_1 \leftrightarrow H_0 + \g$,
 collisions among hydrogen atoms which flip the electron spin
 $H_1 + H_a \leftrightarrow H_0 + H_b$, and  scattering
 of Lyman-$\a$ photons, $H_1 + \g \leftrightarrow H_0 + \g$,
 where  the incident photon excites the electron to the
 { $1p$ or higher levels},
 and in returning to the $1s$ state, the electron
 arrives in a different hyperfine level. Via the decays
 and  absorption of CMB photons, $T_S$
 is attracted to $T_{\rm CMB}$. However,
 at $z\sim 17$, the  kinetic temperature $T_K$ of
 hydrogen is lower than $T_{\rm CMB}$,
 so collisions among hydrogen atoms, and scattering
 of Lyman-$\a$  photons (which tends to pull
 $T_S \to T_K$), try to cool $T_S$.
The timescales for at least some of these processes are
much shorter than the age of the Universe, which during
the matter-dominated era is
$\tau_U \simeq 2 \times 10^8~{\rm years}\,\bigl(\frac{18}{z+1}\bigr)^{3/2}$.
Therefore, the rate to jump
up can be taken equal to the rate to jump down \cite{Furlanetto:2006jb,LM}:
\bea
n_0 \bigl(A_{01} F_{\rm CMB}   +  C_{01} + P_{01} \bigr) =~~~~~~~~~~~ \nonumber  \\
~~~~~~~~~~n_1 \bigl(A_{10} [1+F_{\rm CMB}]   +  C_{10} + P_{10} \bigr),
\label{db}
\eea
where $A$, $C$ and $P$ are respectively the rates
for photon absorption and decays, collisions among hydrogen atoms,
and interactions with Lyman-$\a$ photons.
$F_{\rm CMB}$ is the Bose enhancement factor
due to the phase-space distribution of CMB photons.

The vacuum decay rate of  the upper level $H_1$ is
$A_{10} \simeq 1.9 \times 10^{-30} ~{\rm eV}$,
but the CMB provides a Bose enhancement factor
$F_{\rm CMB}  =1/(e^{E_{10}/T_{\rm CMB}} -1)
\simeq T_{\rm CMB}/E_{10}$.
 So the rate before reionisation is
 \beq
 \label{width}
\Gamma_{10} = A_{10} \frac{T_{\rm CMB}}{E_{10}}
\simeq (1+z)\, 8.5 \times 10^{-29} ~{\rm eV}
\eeq
and the corresponding lifetime
is $\tau_{10} \simeq 2 \times 10^4\frac{18}{1+z}$ years.
The collision and photon-scattering rates can be comparable to
this interaction rate with the CMB,
but can cool the spin temperature.
The photon scattering
terms are more difficult to predict than the collision terms,
because the photons should be produced by the first stars.

\subsection{Axion Dark Matter}

A brief review of the QCD axion can be found in Ref.~\cite{PDB}, or
see {Refs.~\cite{Kim:2008hd,Kuster:2008zz,Irastorza:2018dyq,diCortona:2015ldu}} for more
comprehensive discussions.
We suppose that QCD axions are
the dark matter of our Universe and
that a significant fraction is in
a coherently oscillating background
until $z\sim 5$--10.
This probably implies that the Peccei-Quinn
phase transition occurred before inflation,
because if it occurred afterwards, many
axions might be in miniclusters \cite{mini} that
have already collapsed before $z\sim 17$.

The  classical  axion field, representing the
``realignment axions,'' can be written as \cite{Elmer}
\begin{equation}
a(x,t)=\left[ \frac{VR_0}{R(t)}\right]^{3/2}\!\!
\int \frac{d^3 p}{(2\pi^3)}\,
\frac{\widetilde{a}(\vec{p})\,e^{-ip\cdot x}
+\widetilde{a}^*(\vec{p})\,e^{ip\cdot x}}{\sqrt{2 E_a}},
\label{eqn5}
\end{equation}
where the field is normalised in a comoving
box $V$, and  $R(t)$ is the scale factor
of the Universe, equal to $R_0$ today.
In the non-relativistic limit, where we neglect spatial
gradients of the axion field,
the local energy density
is $\rho_a(x,t)=[\dot a(x,t)^2+m_a^2 a(x,t)^2]/2$ \cite{Elmer}.
  Then the Friedman-Robertson-Walker background density
(spatially averaged density)  is \cite{Elmer}
\bea
\bar{\rho}_a(t) &=&\underset{V\to \infty}{\rm lim}\frac{1}{V}\int_V d^3 x\,\rho_a(x,t)
\nonumber\\[1ex]
&=& m_a  \left[\frac{R_0}{R(t)}\right]^{3} \int  \frac{d^3 p}{(2\pi^3)}\,
 |\widetilde{a}(\vec{p}) |^2.
\label{ps}
\eea
So we see that $|\widetilde{a}(\vec{p}) |^2$ functions as
a phase-space density for the classical axion field
contribution to the average density. 
{(This is  something of a ``trick'', because a classical field  cannot
in general be described by a phase-space distribution, which is part
of the two-point function of two quantum field operators.)}

\subsection{Axion Interaction with Protons and Electrons}

The axion interactions with the electron or proton can be
written \cite{PDB}
\bea
\delta  {\cal L}  =  \frac{C_{e}}{ 2 f_a }(\partial_\mu a)\, \overline{e} \gamma^\mu \g_5 e
+   \frac{C_{p}}{ 2 f_a }(\partial_\mu a)\, \overline{p} \gamma^\mu \g_5 p\,,
\label{L}
\eea
where $f_a$ is the axion decay constant. It is
of order the expectation value of the Peccei-Quinn field,
the complex scalar whose phase is the axion. For QCD axions, one finds
the generic relation \cite{diCortona:2015ldu}
\begin{equation}\label{eq:axionmass}
  m_a=5.70~\mu{\rm eV}\,\left(\frac{10^{12}~{\rm GeV}}{f_a}\right)\,.
\end{equation}
The realignment population of cosmic axions provides a cosmic density of
\cite{PDB}
\begin{equation}
  \Omega_ah^2=0.11\left(\frac{12\,\mu{\rm eV}}{m_a}\right)^{1.19}\Theta_i^2\,,
\end{equation}
where $\Theta_i$ is the initial misalignment angle.
For $m_a=E_{10}=5.9~\mu{\rm eV}$, the requirement for axions to provide all of the dark matter
($\Omega_ah^2=0.11$) corresponds to $\Theta_i=0.66$ and thus is entirely within
the plausible range.

$C_e$ and $C_p$ are model-dependent numerical coefficients.
The hydrogen hyperfine transition depends on
$(C_e-C_p)^2$. In hadronic axion models, notably the
often-cited KSVZ model \cite{K,russes},
one finds
$C_e\simeq 0$, whereas $C_p=-0.47$ \cite{diCortona:2015ldu}
depends only on the axion mixing with the
$\pi^0$, $\eta$ and $\eta'$ mesons
so that $(C_e-C_p)^2=0.22$ is generic.

In non-hadronic models, quarks and leptons carry
Peccei-Quinn charges and
$(C_e-C_p)^2$ becomes model dependent
and potentially vanishes.
One often-cited example
is the DFSZ model \cite{DFS,Z},
where $C_e=\frac{1}{3}\,\sin^2\beta$ and
\smash{$C_p=-0.617+0.435\,\sin^2\beta\pm0.025$} \cite{diCortona:2015ldu} with
$\beta$
describing the ratio of two vacuum expectation values.
We find $(C_e-C_p)^2=0.27$--0.38 and thus never vanishes.

\section{Axion-Hydrogen Interaction}
\label{sec:QFT}

\subsection{Matrix Element}

In order to excite the hyperfine transition,
the axion should give sufficient energy to the H atom and  mediate the
$J= 0 \to 1$ transition of the $1s$ state.
This can occur by the axion
interacting with the spin of either the electron or the proton.
However, to avoid lengthy formulae, we focus on the axion-electron
interaction and discuss how to include the proton
after Eq.~(\ref{Mfi}).

A hydrogen atom of four-momentum $P = m_{\rm H} v$, with $v = (1,\vec{v})$
and the electron in
the $1s$ state, can be written  as a bound state of  a proton $p$
and an electron \cite{P+S,DM}
\begin{widetext}
\begin{equation}\label{bs}
  |H_{ J= 0} (P)\rangle = \sqrt {\frac{2M_{\rm H}}{2m_p 2m_e}}
\int \frac{d^3k}{(2\pi)^3}\,\widetilde{\psi}_{1s}(\vec{k})\,
\frac{|p (-{k} + m_p{v}, -)\rangle \otimes  |e ({k} + m_e{v}, +)\rangle
  - |p (-{k} + m_p{v}, +)\rangle \otimes  |e ({k} + m_e{v}, -)\rangle}{\sqrt{2}},
\end{equation}
\smallskip
\end{widetext}
where $\pm$ labels spin up or down
for the proton and electron, the square-root prefactor is
because states are normalised to $\sqrt{2E}$, and
\smash{$\widetilde{\psi}_{1s}(\vec{k}) = \int d^3 z\,e^{i\vec{k} \cdot  \vec{z}}\psi_{1s}(\vec{z})$}
is the Fourier transform of the electron
wavefunction in the hydrogen $1s$ state,
normalised to $\int d^3 z\,|\psi_{1s}( \vec{z})|^2=1$.

One approach to calculate the S-matrix element for the hydrogen
atom to pass from the $J=0$ to the $J=1$ state
would be to compute the S-matrix element for the transition
${\rm H}_{0} \to {\rm H}_{1}$ in a background axion field,
described as a coherent state.
We have checked that this gives  the same result
as the calculation outlined below,
which  uses  Eq.~(\ref{ps})
as the phase-space distribution for the classical axion
field.
In this second approach, we need the S-matrix element for
$a+ {\rm H}_{0} \to {\rm H}_{1}$
\begin{widetext}
\vskip-6pt
\begin{equation}\label{S}
 {\cal S}_{01}=\int d^4x\,\Bigl\langle  H_{1} (P_1) \Big|
 \frac{C_{e}}{ 2 f_a }\partial_\mu \hat{a}(x) \hat{\overline{e}}(x)
 \gamma^\mu \g_5 \hat{e}(x) \Big|  a(p_a),  H_{0} (P_0) \Bigr\rangle
 =(2\pi)^4\,\delta^4(P_1 - p_a - P_0)\,{\cal M}_{01}\,,
\end{equation}
where operators wear hats.
The rate density is found by integrating
over the initial and final phase-space distributions
\begin{equation}\label{rate}
\gamma_{01}=\int \frac{d^3p_a}{2 E_a(2\pi)^3}
\frac{d^3P_0}{2 E_0(2\pi)^3} \frac{d^3P_1}{2 E_1(2\pi)^3}
F_a F_0 | {\cal M}_{01}|^2\,(2\pi)^4\delta^4(P_1 - p_a - P_0),
\end{equation}
\end{widetext}
where $F_a = |\widetilde{a}(p)|^2$ and $F_0$ are the phase-space
densities of axions and hydrogen atoms.
The interest of calculating the rate
density $\gamma_{01}= n_0 n_a \langle \sigma(a+H_0 \to H_1) v\rangle$,
rather than the cross section $\sigma(a+H_0 \to H_1)$
or the decay rate $\Gamma(H_1 \to H_0+a)$ is that it
avoids dividing then  multiplying by statistical weights
and flux factors.

Using the definition of Eq.~(\ref{bs}), with the
inner product for non-relativistic proton states,
\beq
\langle p (q_1, s_1) |p (q_0, s_0)\rangle =
2 m_p\delta_{s_1 s_0}  (2\pi)^3\delta^3(\vec{q}_1 -\vec{q}_0)
\label{10}
\eeq
and the usual action of field operators on states \cite{P+S}
$\hat{a}(x) |a(p)\rangle = e^{-ip \cdot x}|0\rangle$ and
$\hat{e}(x) |e(k,s)\rangle = e^{-ik \cdot x} u^s(k)|0\rangle$,
{we find
\bea
\kern-1em&&\kern-1em{\cal M}_{01}\times\frac{ 2 f_a m_e }{C_{e} m_{\rm H}}
\nonumber\\
\kern-1em&&{}=
\int\!\frac{d^3 k}{(2 \pi)^3}|\widetilde{\psi}_{1s}(\vec{k})|^2 \langle J=1| p_a^\mu \overline{u}(k)\g_\mu \g_5 u(k)| J=0 \rangle
\nonumber\\
\kern-1em&&{}\approx4 m_e\!\!\int\!\frac{d^3 k}{(2 \pi)^3}|\widetilde{\psi}_{1s}(\vec{k})|^2
\langle 1|  (E_a (\vec{v}_e  - \vec{v}_a)\cdot \vec{S}_e) | 0 \rangle,
\eea
where the second line uses the
non-relativistic expression for
the four-vector inner product{ 
  (neglecting the momentum transfer from axion to electrons) }\cite{CdNP}.
Passing to the hydrogen rest frame, the first term
\smash{($\propto \vec{v}_e \cdot \vec{S}_e $)} disappears because
the expectation value of $\vec{v}_e=\vec{p}_e/m_e$
in the $1s$ state vanishes so that}\footnote{
{Notice that this is different from
  the ``axio-electric effect''
  where the electron changes atomic level \cite{Redondo, PRV, Detal}
  and the rate of relativistic axion emission or absorption,
  relative to the photon rate, is of the order of $(C_e E_a/e f_a)^2$,
  see Eq.~(2.14) of Ref.~\cite{Redondo}. For 
  a pure spin-flip transition, this ratio is of the order of
  $(C_em_e /e f_a)^2$, in our case a factor $(m_e/E_{10})^2\sim 10^{22}$ larger.}}
\beq
 {\cal M}_{01}=\frac{2 m_{\rm H} C_{e}}{  f_a }  \langle J=1|
 \vec{ p}_a \cdot \vec{S_e}  | J=0 \rangle\,.
\label{Mfi}
\eeq
The states $|J \rangle$ are the triplet and singlet
states obtained by combining two spin 1/2 states.
Quantising the spins along $\vec{p}_a$
gives \smash{$\langle J=1|
 \vec{ p}_a \cdot \vec{S}_e  | J=0 \rangle = |\vec{p}_a|/2$}.

When the axion interacts with the spin of
the proton, the calculation  proceeds is a very similar way, with
the electron current replaced by the proton current,
$C_e \to C_p$, and the proton replaced by the electron in Eq.~(\ref{10}).
So including
the axion absorption on the proton and electron gives
a matrix element (in the hydrogen rest frame)
\begin{equation}
    {\cal M}_{01}=
     \frac{2 m_{\rm H} }{  f_a }  \langle J=1|
 \vec{ p}_a \cdot ( C_{e}\vec{S}_e + C_{p}\vec{S}_p ) | J=0 \rangle\,.
\label{Mfiep}
\end{equation}
Quantising the  fermion  spin once more along $\vec{p}_a$
provides $\langle J=1|
 \vec{ p}_a \cdot ( C_{e}\vec{S}_e + C_{p}\vec{S}_p ) | J=0 \rangle =
 (C_e-C_p)\,|\vec{p}_a|/2$.

While it may seem peculiar that protons and electrons contribute on an equal
footing, this arises because the hyperfine transition is  a
pure spin-flip process, and
$C_{e,p}/2 f_a$  can be  of the same order. (In the case of
spin-flip interactions involving the photon, the proton
magnetic moment $g_p e/2m_p$ is much smaller than
that of the electron.)

\subsection{Axion Absorption Rate}
\label{sec:absorption}

In the rest frame of the hydrogen atom,
the rate to jump up the hyperfine splitting
by absorbing an axion is
\bea
\Gamma^{{  (a)}}_{01} &=&
\frac{(C_{e}-C_p)^2}{ 8 m_a f^2_a }\!\! \int \frac{d^3p_a}{(2\pi)^2}
|\vec{p_a}|^2 |\widetilde{a}(\vec{p})|^2 \delta (E_a - E_{10})\nonumber \\
&=&
 \frac{(C_{e}-C_p)^2}{ 8 \pi f^2_a } |\vec{p}_{10}|^3\,|\widetilde{a}(\vec{p}_{10})|^2\,,
\label{estimate}
\eea
where $E_a \sim m_a + |\vec{p}_a|^2/2m_a$
and, assuming $E_{10} > m_a$,
\beq
|\vec{p}_{10}|^2 =2m_a (E_{10}-m_a)\,.
\label{p10}
\eeq
Evaluating this inverse decay rate in the rest frame of the initial hydrogen
atom is justified because the recoil velocity
$|\vec{p}_a|/m_{\rm H} \ll m_a/m_{\rm H} \sim 10^{-15}$
is negligible.
In principle, the rate should be calculated in the co-moving frame of the CMB
by integrating over the momentum distributions of axions
and hydrogen atoms (see Eq.~\ref{rate}).
Instead we will include the hydrogen velocity distribution
by attributing it to the axions.

Some assumption about the axion momentum distribution
is required to get an estimate of the rate
Eq.~(\ref{estimate}).
At very small momenta in the CMB restframe, there are the CDM density
fluctuations\footnote{We suppose that we can still treat
the density fluctuations in the linear approximation.
The axion field should be
smooth on the comoving horizon scale of the
QCD phase transition, because
the axion inherits the density fluctuations from
the radiation bath at the phase transition, and
we suppose that the radiation
bath is homogeneous inside the horizon because
the phase transition is a smooth crossover.
So the maximum axion momentum is
$H_{\rm QCD}$ redshifted until $z=17$.}
which we neglect because we estimate
that they give a
contribution to $E_{10} - m_a$ that is much
smaller than the decay rate~$\Gamma^{{  (a)}}_{10}$.

So in this picture the axions are effectively in the zero-momentum
mode in the CMB frame,
and their effective momentum distribution in the
hydrogen rest frame derives entirely from
hydrogen with kinetic temperature $T_K\lsim 48~{\rm K}$ at $z\sim 17$
as discussed in Sec.~\ref{sec:hyperfine}.  This corresponds to
$\langle v_{\rm H}^2\rangle \lsim 1.33\times10^{-11}$ and thus to a typical
axion momentum $|\vec{p}_a|\sim 4\times10^{-6}\,m_a$ from the
perspective of a typical hydrogen atom.
The corresponding {energy distribution} is much
broader than the atomic line width which is approximately given by
Eq.~(\ref{width}). Therefore, the hydrogen distribution alone assures
us that the effective axion momentum distribution is smooth and
approximately constant on the scale of the line width.

In this case we may simplify the expression for the absorption rate
Eq.~(\ref{rate}) by observing that $|\vec{p}_{10}|^3\, |\widetilde{a}(\vec{p}_{10})|^2$
is dimensionally a number density. If
the  typical  axion momentum
is $ m_a \vec{v}_{\rm H}$, then
the axion number density is $\overline{n}_a =
\overline{\rho}_a/m_a \simeq
m_a^3 |\vec{v}_{\rm H}|^3 |\widetilde{a}|^2/(6\pi)^{3/2}$
(assuming a Maxwell-Boltzmann distribution for the axions,
inherited from the H atoms).
If we further assume that typical axions are on the
hyperfine resonance with $m_a |\vec{v}_{\rm H}|\simeq |\vec{p}_{10}|$,
then the absorption rate is
\begin{equation}
\Gamma^{{  (a)}}_{01}\simeq
 \frac{(C_{e}-C_p)^2}{8 \pi f^2_a} \frac{(6\pi)^{3/2}\,\overline{\rho}_a}{m_a}\,.
\end{equation}
If axions are the dark matter, $\overline{\rho}_a=\Omega_{\rm DM}\rho_{\rm crit}(1+z)^3
=5.2\times10^{-8}~{\rm eV}^4\,(\frac{1+z}{18})^3$. With $m_a=E_{10}=5.9~\mu{\rm eV}$
and the corresponding $f_a=0.97\times10^{12}~{\rm GeV}$ from
Eq.~(\ref{eq:axionmass}) we find what is essentially an upper limit to
the absorption rate at $z\sim17$
\begin{equation}\label{est2b}
  \Gamma^{{  (a)}}_{01}\alt (C_p-C_e)^2~~3\times10^{-44}~{\rm eV}\,.
\end{equation}
This rate is about 17 orders of magnitude smaller than the standard
rates {  involving photons }that are of the order of Eq.~(\ref{width}).

If the effective axion momentum distribution is due entirely to the hydrogen
kinetic temperature, this rate requires $m_a$ being tuned to
$E_{10}$ within more than ten figures.
A much larger transition rate could be obtained only
if the dark matter axions are taken to have
momenta in a narrow range $\Delta p$ around
$|\vec{p}_{10}|$. Then the number density is
$\propto\Delta p |\vec{p}_{10}|^2\,|\widetilde{a}(\vec{p}_{10},t)|^2$,
and the rate estimate of Eq.~(\ref{est2b}) would be enhanced
by a factor $|\vec{p}_{10}|/\Delta p$.
Whether this is reasonable will depend on the
origin of the axion phase-space distribution.

\section{Calculating in Quantum Mechanics}
\label{sec:QM}

In their chapter 13, Cohen-Tannoudji
{\it et al.} \cite{CT} calculate, in quantum
mechanics, the rate at which a
classical electromagnetic wave,
with frequency tuned to the  energy
difference between two atomic levels,
can excite the upper level.
This calculation is analogous to our problem,
because  the magnetic part of the electromagnetic
wave interacts with the electron spin,
similarly to the axion.

Using  time-dependent perturbation theory,
in the presence of an
oscillating background electromagnetic field,
Cohen-Tannoudji {\it et al.} calculate a
transition matrix element between levels  of
the hydrogen atom. We
include the axion field in the Hamiltonian of the electron as
 \beq
\frac{2C_e}{f_a}\,a(t)\, \cos(E_a t)\, \vec{p}_a \cdot \vec{S}_e\,,
 \label{W}
\eeq
where the time dependence
of $a(t)$ encodes cosmic redshifting.
Adapting the Cohen-Tannoudji {\it et al.} calculation gives
 a matrix element $ a(t) \times$ the result
 given in Eq.~(\ref{Mfi}).  The extra factor
 of axion field is expected in the background
 field approximation.  In the
 electromagnetic calculation, there
 is a term proportional to the momentum
 of the electron, which we do not include
 because it vanishes for $1s$ hyperfine transitions
 (as discussed
 after Eq.~\ref{Mfiep}).
This gives a transition  probability
\beq
{\cal P}_{0 \to 1}
 = \frac{C_e^2}{2 f_a^2} |\vec{p}|^2 {a}^2(t)   \pi  \delta(E_a - E_{10})\,t\,.
\eeq
Cohen-Tannoudji {\it et al.} then observe that
incoming radiation usually has a spectrum,
and introduce   ${\cal I}(\omega)d \omega$, the flux  per unit frequency
$\omega$, such that  they integrate
over the $\delta$-function. So in order
to check our calculation against their result, we need  to find the
appropriate analogy of  ${\cal I}(\omega)$ for the
non-relativistic axion.
If  we  replace
\beq
|\vec{p}|^2 {a}^2 (t) 
\to \int\frac{d^3 p}{2 E_a(2\pi)^3} |\vec{p}|^2 |\widetilde{a}(p)|^2
\eeq
this has the correct dimensions,  allows us to
integrate over the momentum distribution
of the initial-state axions (three-momentum is
a more useful label than frequency for the
non-relativistic axion), and
gives the same rate as obtained in  the previous section
(after retrieving the redshift factor
scaled out of Eq.~\ref{eqn5}).

\section{Discussion}
\label{sec:disc}

We have estimated the hydrogen hyperfine transition rate
$H_0 \leftrightarrow H_1$
induced by the absorption of a dark matter
axion whose mass $m_a$ was
comparable to the hyperfine splitting
$E_{10}=5.9~\mu{\rm eV}$. We have found, in Eq.~(\ref{est2b}),
a rate  that is negligibly small: the
upper hyperfine level H$_1$  takes longer
than the age of the Universe to jump down while emitting an axion.
Indeed, it is $\sim 10^{-17}$ times smaller
than the  rate to emit a photon.

As expected, our rate estimate
$(C_e - C_p)^2 n_a/f_a^2$
is proportional  to the axion density
and to the square of the coupling constants of Eq.~(\ref{L}).
The calculation was performed with some formal detail,
and in three different ways, because the estimates one
could make from the literature diverge widely.
As  mentioned in footnote~2,  a naive
extrapolation of the ``axio-electric rate''
would underestimate the rate by 22 orders of magnitude.
On the other hand,  Lambiase and Mohanty \cite{LM}
found an additional enhancement factor
of $m_a/\Gamma_{10}$, where $\Gamma_{10}$
is the decay rate $H_1 \to \g + H_0$ (see
Eq.~\ref{width}).
This 
rate could be fast enough to change the spin temperature
of hydrogen prior to reionisation.
According to our calculation, such an enhancement factor
could arise from a
very narrow axion phase-space distribution,
of width $\Delta p = \Gamma_{10}$, as explained at the end of
Sec.~\ref{sec:absorption}.
But the effective axion phase-space distribution
should encode the thermal hydrogen distribution, so we do
not think this enhancement mechanism is reasonable.

We also observe
that the effect of axions is in the direction of
heating, not cooling, the hydrogen spin temperature.
The main
effect derives from absorbing or emitting an axion, so
the very small axion kinetic temperature does not imply
cooling.
This is most easily seen from a rate equation of
the form of Eq.~(\ref{db}) for axions alone,
$n_0 X_{01} F_a = n_1  X_{10}(1+F_a)$,
 where $\Gamma^{{  (a)}}_{01}$ of Eq.~(\ref{estimate}) is
 $X_{01}F_a$, and $F_a$ is  the axion phase-space density.
 In the  limit of large occupation numbers,
 where $1+ F_a\simeq F_a$, detailed
balance implies
$ n_1/n_0  \to  3$ so that $T_S\to\infty$.

In principle one could derive a limit on axions
with masses around $E_{10}$ from the requirement to
avoid excessive heating of the spin temperature.
However, our rate estimate is far too small to use the effect
in this way, and appears to only
be able to exclude axions whose mass matches
the hyperfine splitting to ten significant figures.

In summary, we have considered the prospects for
  QCD axions, which compose the dark matter
  of our Universe and have a mass of order
  the hyperfine spitting, to affect the hydrogen
  spin temperature in the epoch prior to reionisation.
  We have presented  two calculations
 of the rate for axions to induce hyperfine-flip
transitions:
in quantum mechanics with a background axion field,
and in quantum field theory with a phase-space density.
Even in the most optimistic case, our rate estimate
is some 17 orders of magnitude too small to have
a tangible impact.

\section*{Acknowledgements}

AA  thanks the LUPM laboratory
at the University of Montpellier for
their welcome during  this summer
project in the Masters programme
``Cosmos, Champs et Particules''.
GR acknowledges partial support by the Deutsche Forschungsgemeinschaft
through Grants No.\  EXC 153 (Excellence Cluster ``Universe''), SFB 1258
(Collaborative Research Center ``Neutrinos, Dark Matter, Messengers''),
and by the European Union through Grant
No.\ H2020-MSCA-ITN-2015/674896 (Innovative Training Network
``Elusives'')


\begin{thebibliography}{00}

\bibitem{EDGE}
  J.~D.~Bowman, A.~E.~E.~Rogers, R.~A.~Monsalve, T.~J.~Mozdzen and
  N.~Mahesh,
  ``An absorption profile centred at 78~megahertz in the
  sky-averaged spectrum,''
  Nature {\bf 555}, 67 (2018).

\bibitem{Tashiro:2014tsa}
  H.~Tashiro, K.~Kadota and J.~Silk,
  ``Effects of dark matter-baryon scattering on redshifted 21 cm signals,''
  Phys.\ Rev.\ D {\bf 90}, 083522 (2014)
  [arXiv:1408.2571].

\bibitem{Munoz:2015bca}
  J.~B.~Mu{\~n}oz, E.~D.~Kovetz and Y.~Ali-Haïmoud,
  ``Heating of baryons due to scattering with dark matter during the Dark Ages,''
  Phys.\ Rev.\ D {\bf 92}, 083528 (2015)
  [arXiv:1509.00029].

\bibitem{Barkana:2018lgd}
  R.~Barkana,
  ``Possible interaction between baryons and dark-matter particles
  revealed by the first stars,''
  Nature {\bf 555}, 71 (2018)
  [arXiv:1803.06698].

\bibitem{Fialkov:2018xre}
  A.~Fialkov, R.~Barkana and A.~Cohen,
  ``Constraining baryon--dark matter scattering with the
   cosmic dawn 21-cm signal,''
   Phys.\ Rev.\ Lett.\  {\bf 121}, 011101 (2018)
  [arXiv:1802.10577].

\bibitem{Barkana:2018qrx}
 R.~Barkana, N.~J.~Outmezguine, D.~Redigolo and T.~Volansky,
  ``Strong constraints on light dark matter interpretation of the EDGES signal,''
  Phys.\ Rev.\ D {\bf 98} (2018) no.10,  103005
  doi:10.1103/PhysRevD.98.103005
  [arXiv:1803.03091 [hep-ph]].

\bibitem{Berlin:2018sjs}
  A.~Berlin, D.~Hooper, G.~Krnjaic and S.~D.~McDermott,
  ``Severely constraining dark matter interpretations of the 21-cm anomaly,''
  Phys.\ Rev.\ Lett.\  {\bf 121}, 011102 (2018)
  [arXiv:1803.02804].

\bibitem{LM}
  G.~Lambiase and S.~Mohanty,
  ``The 21-cm axion,''
  arXiv: 1804.05318.

\bibitem{gravtherm1}
 N.~Houston, C.~Li, T.~Li, Q.~Yang and X.~Zhang,
  Phys.\ Rev.\ Lett.\  {\bf 121} (2018) no.11,  111301
  doi:10.1103/PhysRevLett.121.111301
  [arXiv:1805.04426 [hep-ph]].
  
\bibitem{gravtherm2}
  P.~Sikivie,
  ``Axion dark matter and the 21-cm signal,''
  arXiv:1805.05577.

\bibitem{otheraxion}
  K.~Lawson and A.~R.~Zhitnitsky,
  ``The 21~cm absorption line and axion quark nugget dark matter model,''
  arXiv:1804.07340.

\bibitem{Feng:2018rje}
  C.~Feng and G.~Holder,
  ``Enhanced global signal of neutral hydrogen due to excess radiation at cosmic dawn,''
  Astrophys.\ J.\  {\bf 858}, L17 (2018)
  [arXiv:1802.07432].

\bibitem{Ewall-Wice:2018bzf}
A.~Ewall-Wice, T.-C.~Chang, J.~Lazio, O.~Dore, M.~Seiffert and R.~A.~Monsalve,
  ``Modeling the Radio Background from the First Black Holes at Cosmic Dawn: Implications for the 21 cm Absorption Amplitude,''
  Astrophys.\ J.\  {\bf 868} (2018) no.1,  63
  doi:10.3847/1538-4357/aae51d
  [arXiv:1803.01815 [astro-ph.CO]].
  

\bibitem{PPRU}
  M.~Pospelov, J.~Pradler, J.~T.~Ruderman and A.~Urbano,
  ``New physics in the Rayleigh-Jeans tail of the CMB,''
  Phys.\ Rev.\ Lett.\  {\bf 121}, 031103 (2018)
  [arXiv:1803.07048].

\bibitem{Moroi}
  T.~Moroi, K.~Nakayama and Y.~Tang,
  ``Axion-photon conversion and effects on 21~cm observation,''
  Phys.\ Lett.\ B {\bf 783}, 301 (2018)
  [arXiv:1804.10378].

\bibitem{FalkowskiPetraki}
  A.~Falkowski and K.~Petraki,
  ``21~cm absorption signal from charge sequestration,''
  arXiv:1803.10096.

\bibitem{DMconstraints1}
 S.~Fraser {\it et al.},
  ``The EDGES 21 cm Anomaly and Properties of Dark Matter,''
  Phys.\ Lett.\ B {\bf 785} (2018) 159
  doi:10.1016/j.physletb.2018.08.035
  [arXiv:1803.03245 [hep-ph]].


\bibitem{DMconstraints2}
 G.~D'Amico, P.~Panci and A.~Strumia,
  ``Bounds on dark matter annihilations from 21 cm data,''
  Phys.\ Rev.\ Lett.\  {\bf 121}, 011103 (2018)
  [arXiv:1803.03629].

\bibitem{DMconstraints3}
J.~B.~Muñoz, C.~Dvorkin and A.~Loeb,
  ``21-cm Fluctuations from Charged Dark Matter,''
  Phys.\ Rev.\ Lett.\  {\bf 121} (2018) no.12,  121301
  doi:10.1103/PhysRevLett.121.121301
  [arXiv:1804.01092 [astro-ph.CO]].


\bibitem{DMconstraints4}
K.~Cheung, J.~L.~Kuo, K.~W.~Ng and Y.~L.~S.~Tsai,
  ``The impact of EDGES 21-cm data on dark matter interactions,''
  Phys.\ Lett.\ B {\bf 789} (2019) 137
  doi:10.1016/j.physletb.2018.11.058
  [arXiv:1803.09398 [astro-ph.CO]].


\bibitem{DMconstraints5}
  H.~Liu and T.~R.~Slatyer,
  ``Implications of a 21-cm signal for dark matter annihilation and decay,''
  Phys.\ Rev.\ D {\bf 98}, 023501 (2018)
  [arXiv:1803.09739].

\bibitem{DMconstraints6}
T.~R.~Slatyer and C.~L.~Wu,
  ``Early-universe constraints on dark matter-baryon scattering and their
  implications for a global 21~cm signal,''
  Phys.\ Rev.\ D {\bf 98}, 023013 (2018)
  [arXiv:1803.09734].

\bibitem{DMconstraints7}
 S.~Clark, B.~Dutta, Y.~Gao, Y.~Z.~Ma and L.~E.~Strigari,
  ``21 cm limits on decaying dark matter and primordial black holes,''
  Phys.\ Rev.\ D {\bf 98} (2018) no.4,  043006
  doi:10.1103/PhysRevD.98.043006
  [arXiv:1803.09390 [astro-ph.HE]].
  

\bibitem{Mirizzi}
  C.~Evoli, M.~Leo, A.~Mirizzi and D.~Montanino,
  ``Reionization during the dark ages from a cosmic axion background,''
  JCAP {\bf 1605}, 006 (2016)
  [arXiv:1602.08433].

\bibitem{Sikivie:2014lha}
  P.~Sikivie,
  ``Axion dark matter detection using atomic transitions,''
  Phys.\ Rev.\ Lett.\  {\bf 113}, 201301 (2014)
  [arXiv: 1409.2806].

\bibitem{Redondo}
  J.~Redondo,
  ``Solar axion flux from the axion-electron coupling,''
  JCAP {\bf 1312} (2013) 008
  [arXiv:1310.0823].

\bibitem{PRV}
  M.~Pospelov, A.~Ritz and M.~B.~Voloshin,
  ``Bosonic super-WIMPs as keV-scale dark matter,''
  Phys.\ Rev.\ D {\bf 78}, 115012 (2008)
  [arXiv:0807.3279].

 \bibitem{Detal}
  A.~Derevianko, V.~A.~Dzuba, V.~V.~Flambaum and M.~Pospelov,
  ``Axio-electric effect,''
  Phys.\ Rev.\ D {\bf 82}, 065006 (2010)
  [arXiv:1007.1833].

\bibitem{CT}
  C.~Cohen-Tannoudji, B.~Diu, F.~Lalo\"e, {\em M\'ecanique Quantique}, Tome II,
  chapter 12 (Hermann, 1997).

\bibitem{BjD}
  J.~D.~Bjorken and S.~D.~Drell,
  {\em Relativistic Quantum Mechanics}, chapter 4.4
  (McGraw-Hill, 1965)

\bibitem{PDB}
  M.~Tanabashi {\em et al}. (Particle Data Group),
  ``Review of particle physics (2018),''
  Phys. Rev. D {\bf 98}, 030001 (2018).

\bibitem{Kim:2008hd}
  J.~E.~Kim and G.~Carosi,
  ``Axions and the strong CP problem,''
  Rev.\ Mod.\ Phys.\  {\bf 82}, 557 (2010)
  [arXiv: 0807.3125].

\bibitem{Kuster:2008zz}
  M.~Kuster, G.~Raffelt and B.~Beltr\'an (eds.),
  {\em Axions: Theory, cosmology, and experimental searches},
   Lect.\ Notes Phys.\  {\bf 741}, 1-- 258 (2008).

\bibitem{Irastorza:2018dyq}
  I.~G.~Irastorza and J.~Redondo,
  ``New experimental approaches in the search for axion-like particles,''
  Prog.\ Part.\ Nucl.\ Phys.\  {\bf 102}, 89 (2018)
  [arXiv:1801.08127].

\bibitem{diCortona:2015ldu}
  G.~Grilli di Cortona, E.~Hardy, J.~Pardo Vega and G.~Villadoro,
  ``The QCD axion, precisely,''
  JHEP {\bf 1601}, 034 (2016)
  [arXiv:1511.02867].

\bibitem{mini}
  C.~J.~Hogan and M.~J.~Rees,
  ``Axion miniclusters,''
  Phys.\ Lett.\ B {\bf 205}, 228 (1988).

\bibitem{Elmer}
  S.~Davidson and M.~Elmer,
  ``Bose Einstein condensation of the classical axion field in cosmology?,''
  JCAP {\bf 1312} (2013) 034
  [arXiv:1307.8024].


\bibitem{Pit}
  F.~Dalfovo, S.~Giorgini, L.~P.~Pitaevskii and S.~Stringari,
  ``Theory of Bose-Einstein condensation in trapped gases,''
  Rev.\ Mod.\ Phys.\ {\bf 71}, 463 (1999).

\bibitem{K}
  J.~E.~Kim,
  ``Weak interaction singlet and strong CP Invariance,''
  Phys.\ Rev.\ Lett.\  {\bf 43}, 103 (1979).

\bibitem{russes}
  M.~A.~Shifman, A.~I.~Vainshtein and V.~I.~Zakharov,
  ``Can confinement ensure natural CP invariance of strong interactions?,''
  Nucl.\ Phys.\ B {\bf 166}, 493 (1980).

\bibitem{DFS}
  M.~Dine, W.~Fischler and M.~Srednicki,
  ``A simple solution to the strong CP problem with a harmless axion,''
  Phys.\ Lett.\ B {\bf 104}, 199 (1981).

\bibitem{Z}
  A.~R.~Zhitnitsky,
  ``On possible suppression of the axion hadron interactions,''
  Sov.\ J.\ Nucl.\ Phys.\ {\bf 31}, 260 (1980)
  [Yad.\ Fiz.\  {\bf 31}, 497 (1980)].

\bibitem{Furlanetto:2006jb}
  S.~Furlanetto, S.~P.~Oh and F.~Briggs,
  ``Cosmology at low frequencies: The 21 cm transition and the high-redshift universe,''
  Phys.\ Rept.\ {\bf 433}, 181 (2006)
  [astro-ph/0608032].

\bibitem{P+S}
  M.~E.~Peskin and D.~V.~Schroeder,
  {\em An Introduction to quantum field theory}
  (Addison-Wesley, 1995).

\bibitem{DM}
The bound state formalism is given in appendix B of:\\
A.~L.~Fitzpatrick, W.~Haxton, E.~Katz, N.~Lubbers and Y.~Xu,
  ``The effective field theory of dark matter direct detection,''
  JCAP {\bf 1302}, 004 (2013)
  [arXiv:1203.3542].

\bibitem{CdNP}
  M.~Cirelli, E.~Del Nobile and P.~Panci,
  ``Tools for model-independent bounds in direct dark matter searches,''
  JCAP {\bf 1310}, 019 (2013)
  [arXiv:1307.5955].

\end{thebibliography}
\end{document}